# RADIFUSION: A multi-radiomics deep learning based breast cancer risk prediction model using sequential mammographic images with image attention and bilateral asymmetry refinement

Hong Hui Yeoh[1], Andrea Liew[1], Raphaël Phan[7], Fredrik Strand[3,4], Kartini Rahmat[5], Tuong Linh Nguyen[6], John L. Hopper[6], Maxine Tan[1,2,*]

[1] Electrical and Computer Systems Engineering Discipline, School of Engineering, Monash University Malaysia, Bandar Sunway 47500, Malaysia.

[2]School of Electrical and Computer Engineering, The University of Oklahoma, Norman, OK 73019, USA.

[3]Department of Oncology and Pathology, Karolinska Institute, Solna, Sweden.

[4]Breast Radiology, Karolinska University Hospital, Solna, Sweden.

[5]Department of Biomedical Imaging and University of Malaya Research Imaging Centre, Faculty of Medicine, University of Malaya, 50603 Kuala Lumpur, Malaysia.

[6]Centre for Epidemiology & Biostatistics, The University of Melbourne School of Population and Global Health, Melbourne, Victoria, Australia.

[7] School of Information Technology, Monash University Malaysia, Bandar Sunway 47500, Malaysia.

[*]Corresponding author. Email address: Maxine.Tan@monash.edu (Maxine Tan)

## ABSTRACT

Breast cancer is a significant public health concern and early detection is critical for triaging high risk patients. Sequential screening mammograms can provide important spatiotemporal information about changes in breast tissue over time. In this study, we propose a deep learning architecture called RADIFUSION that utilizes sequential mammograms and incorporates a linear image attention mechanism, radiomic features, a new gating mechanism to combine different mammographic views, and bilateral asymmetry-based finetuning for breast cancer risk assessment. We evaluate our model on a screening dataset called Cohort of Screen-Aged Women (CSAW) consisting of 8,723 patients altogether. Based on results obtained on the independent testing set consisting of 1,749 women, our approach achieved superior performance compared to other state-of-the-art models with area under the receiver operating characteristic curves (AUCs) of 0.905, 0.872 and 0.866 in the three respective metrics of 1-year AUC, 2-year AUC and > 2-year AUC. Our study highlights the importance of incorporating various deep learning mechanisms, such as image attention, radiomic features, gating mechanism, and bilateral asymmetry-based fine-tuning, to improve the accuracy of breast cancer risk assessment. We also demonstrate that our model's performance was enhanced by leveraging spatiotemporal information from sequential mammograms. Our findings suggest that RADIFUSION can provide clinicians with a powerful tool for breast cancer risk assessment.

## 1. INTRODUCTION

Breast cancer is the most commonly diagnosed cancer among women patients and is the second leading cause of cancer related mortality in the US [1]. Breast cancer alone accounts for 30% of all female cancers and accounts for the 77% increase in cancer risk during early adulthood of 20 – 49 years old for the female population. Early detection of breast cancer is crucial as it greatly improves the chances of successful treatment and patient survivability. As such, regular screening tests, such as mammograms and clinical breast exams can help to detect breast cancer in its early stages.

Besides early detection, risk assessment is also important in the prevention and management of breast cancer. This involves identifying and evaluating various factors that may increase an individual's risk of developing the disease, such as age, family history, genetics, and lifestyle factors. Early risk models such as the Gail model [2] and Tyrer-Cuzick model [3] leverage such factors, though having modest performance. Breast density and parenchymal texture patterns are also factors that have been linked to an increased risk of breast cancer and their assessment have also proven useful for risk prediction [4-6]. However, we believe a deep learning model can also capture contextual information such as breast density or parenchymal texture patterns at the same time from the mammograms. Deep learning has been implemented as a solution to develop accurate breast cancer risk assessment models. Image-based deep learning models [7-9] have shown very promising results and suggest that deep learning has the ability to recognize more predictive information in mammograms than traditional risk models.

To the best of our knowledge, all the deep learning based studies with the exception of one study [10] have not taken into consideration information provided by previous mammographic screenings in their models. In standard clinical practice, women undergo multiple screenings which are used to monitor the changes of dense breast tissue and appearance over time. Longitudinal studies of mammograms have been proposed in other tasks [11-13] and have shown that incorporating previous screenings can improve a model's predictive capability. A new study that proposed using gated recurrent units (GRUs) [10] for breast cancer risk also showed similar improvements, though they did not consider radiomic features nor any attention mechanism.

Although many studies have been proposed to predict cancer risk using clinical risk factors [2, 3], breast parenchymal texture analysis/radiomics based methods [14], deep learning on the current year mammogram [6], or bilateral asymmetry [15] based methods, these studies were all individualized models and did not combine all factors to develop a comprehensive risk model. On the other hand, all aspects of radiomics, bilateral asymmetry between left and right breasts (as cancer typically develops in only one breast and normal breasts typically have symmetrical appearances), features learnt by deep learning and clinical risk factors (e.g., age, BMI, family history, etc.) have all shown to provide useful information *individually* for risk prediction. What is lacking in the literature is a risk model that combines all these factors into one comprehensive model.

The Non-Local network was first introduced for video classification [16] and its purpose is to increase the receptive field of the Convolutional Neural Network (CNN) to the input image size by considering a weighted sum of all generated features in the whole video. In high resolution mammography images, a self-attention based method like Non-Local will be very useful to extract subtle patterns/tissues in the breast that indicate the presence of breast cancer risk. Another advantage of Non-Local networks over other Transformer based methods like VIT [17] is that Non-Local networks can be used for 3D video. The significance of this to our work is that we can model the temporal series of mammograms as a 3D video. In this way, cancer risk can be detected from a series of sequential mammograms of the same patient (to detect subtle changes in the patients' breast tissue), similar to what is done by radiologists in clinical practice. However, the downside of the Transformer/self-attention model is that its complexity is quadratic to the input sequence length, as the dot-product between the input representations at each pair of positions needs to be computed [18]. It is thus extremely costly for standard Transformer/self-attention models to efficiently handle long input sequences, let alone large 2D images (mammograms) and 3D videos.

In this study, we propose a new deep learning architecture called RADIFUSION, capable of breast cancer risk prediction by modelling the changes in breast tissue structure over time. Our proposed deep learning model incorporates novel methods and combines/augments the previously studied risk factors including deep learning features, radiomics, bilateral asymmetry feature detection, combining features from different craniocaudal (CC) and mediolateral oblique (MLO) views, etc. into one comprehensive risk model. First, we present a novel attention mechanism that can efficiently handle increased demands in computation and memory in 3D sequential screening mammograms. To the best of our knowledge, this is the first additive attention model for images/videos. This new attention mechanism enables the model to focus on the most relevant features throughout multiple prior screenings of a patient in a fast and accurate manner. Second, we present new Radiomics features that complement the deep learning features in enhancing the model's capability for risk prediction. Third, we propose a gating mechanism that can automatically determine the optimal weights/attention that should be given to the (CC or MLO) mammographic views to enhance the network's performance. Lastly, we present a new pseudo labelling method that uses information extracted from bilateral asymmetry to further finetune and improve the performance of our model.

It is the authors' hope that our new model would be able to assist radiologists in triaging high-risk patients from low-risk patients. We also hope that our new model can assist radiologists in making individualized screening recommendations so that women who have a higher predisposition to cancer in the short-term (e.g., 1 to 2 years) are screened frequently compared to those with lower short-term risk. We examine the performance of our models on a comprehensive screening dataset, namely the Cohort of Screen-Aged Women (CSAW) dataset [19].

The outcomes of this research are:

1. A novel deep learning architecture called RADIFUSION that can capture changes in breast tissue over spacetime through sequential mammographic screenings.
2. Big data method of combining deep learning features with radiomic features and combining features from craniocaudal (CC) and mediolateral oblique (MLO) views.
3. A novel attention block called SHIFT (Spatial Channel Image Fastformer) that can efficiently capture both spatial and channel attention across multiple mammographic screenings.
4. A new multi-view gating mechanism to optimally weigh information from craniocaudal (CC) and mediolateral oblique (MLO) views.
5. A new Bilateral Symmetry-aware finetuning method to further refine the model using bilateral asymmetry risk score between contralateral breasts.
6. Validation of our new risk model and its performance on a comprehensive screening dataset.

## 2. MATERIALS & METHODS

### *2.1 Dataset*

We analyse our new method on a comprehensive screening dataset called the Cohort of Screen-Aged Women (CSAW) dataset [19]. For the CSAW dataset, the ethical review board of Stockholm permitted the research study and waived the requirement for individual informed consent (EPN 2016/2600-31), with additions approved by the Ethical Review Authority of Sweden (EPM 2019-01946, EPM 2019-03638, EPM 2021-01030).

The acquired dataset is a subset of CSAW [19], which is a complete population-based cohort of women 40 to 74 years of age invited for screening in the Stockholm region, Sweden, between 2008 and 2016. Mammograms of multiple views were collected every 18 to 24 months from women aged 40 to 74. All images were acquired on Hologic mammography equipment. Patients with examinations that did not include all four standard views and were outside the screening age range of 40-74 were excluded.

The provided dataset consists of 8,723 patients with 7,850 controls and 873 cases, altogether. For the 873 cases, two key clinical risk factors were censored in the dataset, namely the patients' age and their days from screening to cancer diagnosis. For the patients' age, the category/value **1** was given to patients at **40-55** years of age at mammography, and **2** to patients at **55+** years of age at mammography. For patients that were diagnosed with cancer, the category **1** was given to patients diagnosed with cancer at **<= 60** days from screening, **2** to **61-729** days, and **3** to **730+** days. This puts each cancer patient into broad categories for anonymization purposes, which might limit what the model can learn from the clinical risk factors. Of the 873 cases in these 3 broad categories, some had sequential mammograms up to the current year (up to 5), whereas others had only one screening mammogram when cancer was detected by the radiologist. The breakdown of the 873 cases in Categories 1 to 3, in terms of number of available sequential screenings per patient is tabulated in Table 1 and shown in Fig. 1. An example of a cancer patient with three longitudinal prior mammogram examinations in our dataset is shown in Fig. 2.

Table 1. Breakdown of CSAW dataset by patient age and days from screening to cancer diagnosis

| | Subgroup | Positive | Negative |
|---|---|---|---|
| Age | 40 - 55 | 319 (37%) | 3853 (49%) |
| | 55+ | 554 (63%) | 3997 (51%) |
| | | | |
| Cancer diagnosis | 0 | - | 7850 |
| | 1 (<= 60 days) | 524 (60%) | - |
| | 2 (61-729 days) | 217 (25%) | - |
| | 3 (730+ days) | 132 (15%) | - |

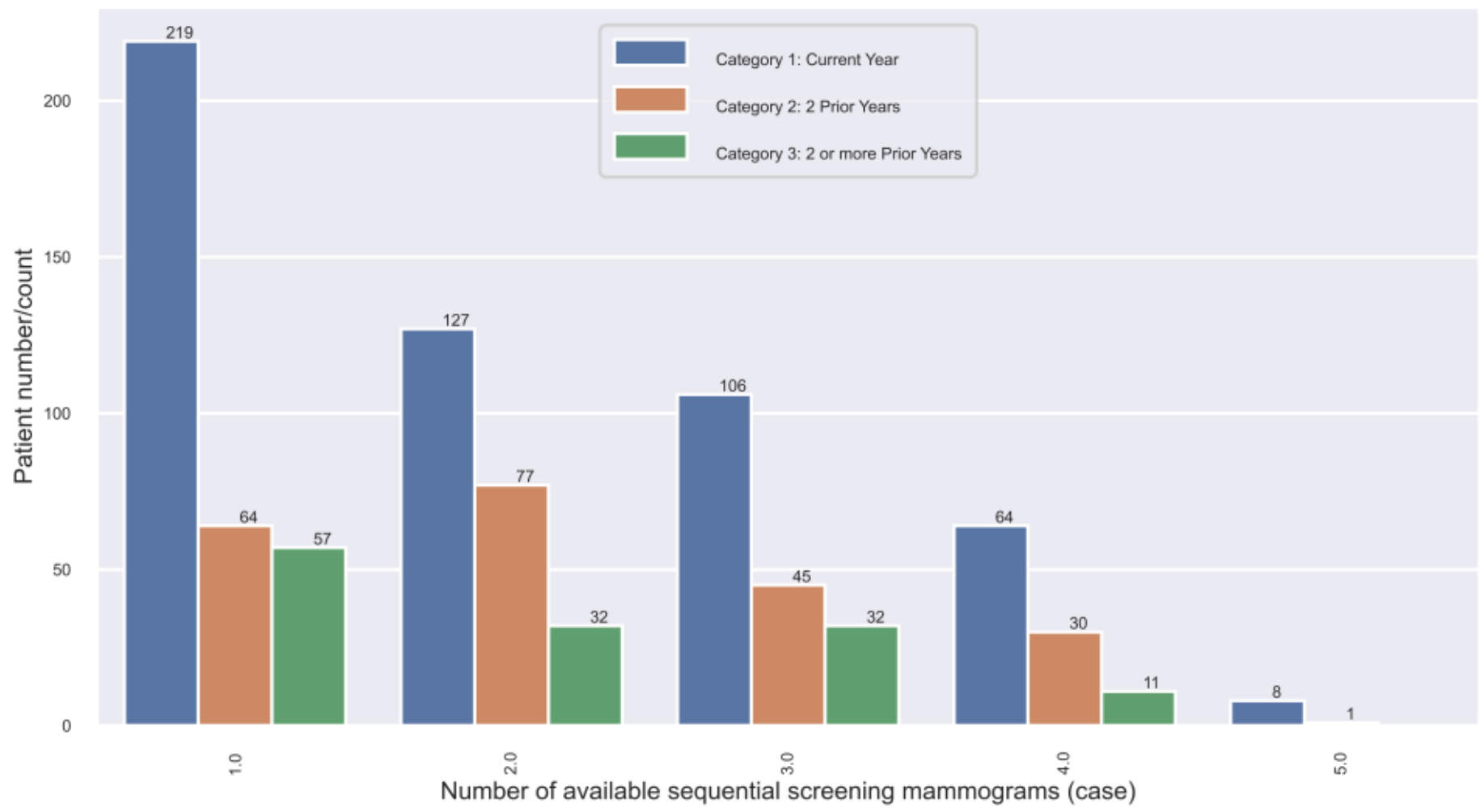

Fig. 1. Number of available sequential screenings per patient within three current and prior year cancer diagnosis categories in the CSAW dataset.

*2.2 Data Preprocessing*

We applied some pre-processing steps to the mammograms before feeding them to our risk model. First, we segmented and removed unrelated labels and the background through Otsu's thresholding. All the mammographic views were resized to 256 by 256 pixels, with zero padding applied to maintain the image's aspect ratio. We also normalized the images to have zero mean and unit variance by calculating the pixel mean and standard deviation across the training set of each cross-validation fold.

To utilize previous mammography screening images, we construct the 3D representation of the mammogram, where processed mammograms are stacked together in order of the most recent screenings. This can be interpreted as a video form of the mammograms, and different screenings account for different timepoints in the video. For patients with insufficient screenings, we duplicated their most recent screenings to fill the timepoint gaps in the video.

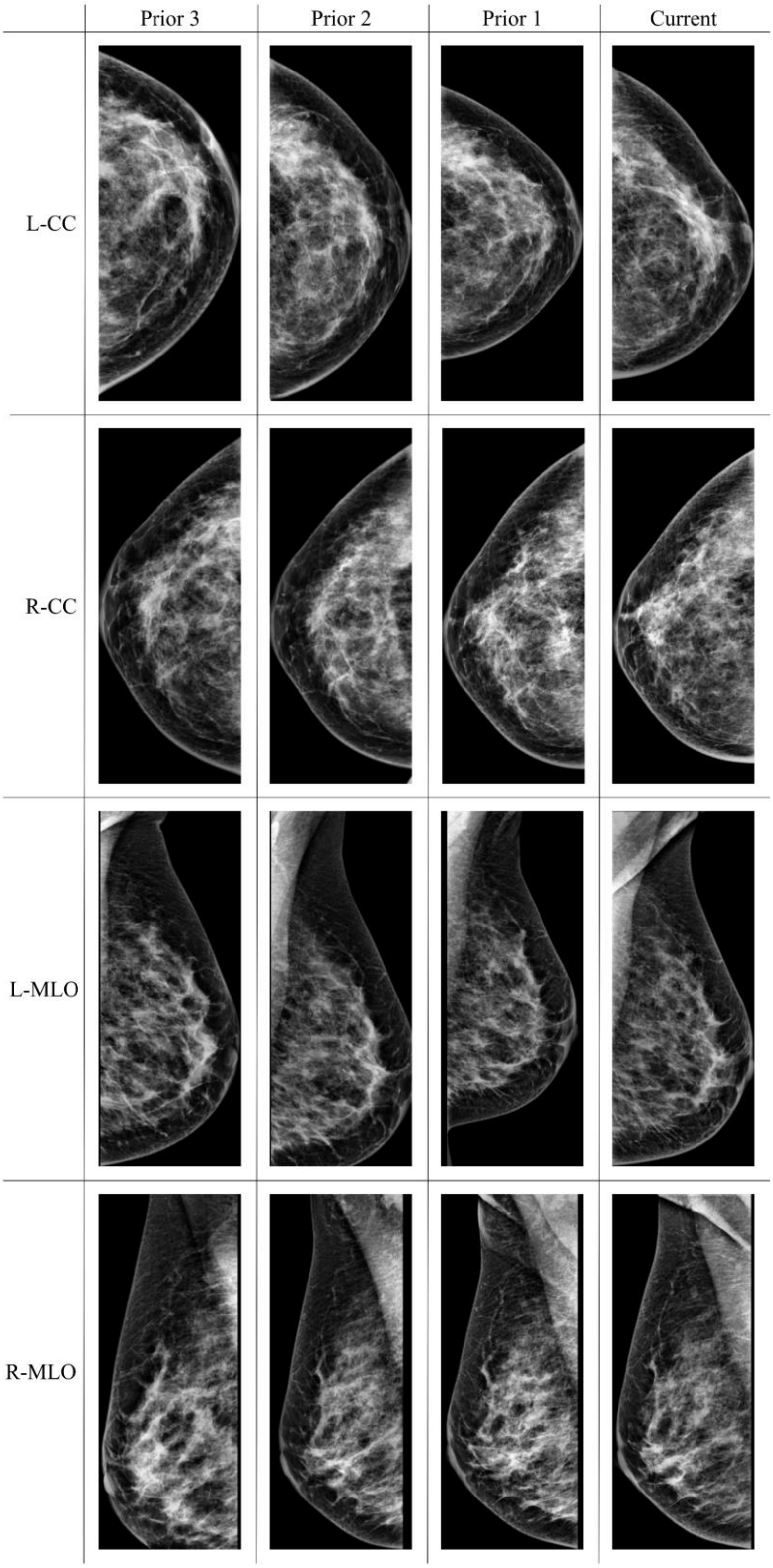

Fig. 2. An example of three longitudinal prior mammogram examinations of a cancer patient in our dataset. The prior examinations, labelled as prior-1 through prior-3, were collected from 2009 to the current one at 2013. Each prior examination consists of four images, consisting of CC and MLO views for both left and right breasts.

*2.3 Three-dimensional Convolutional Neural Network (CNN) Architecture*

Resnet18 was used as the backbone of our CNN model. It serves as our feature extractor which compresses the image representation into a vector with dimensionality of 512 as shown in Fig. 5. The base architecture, which was designed for image classification was extended to a three-dimensional (3D) model by "inflating" the kernels [16, 20].

Our approach is adopted from [16], whereby a 2D $k \times k$ kernel can be inflated to a 3D $t \times k \times k$ kernel that spreads across $t$ frames. The weights from the pretrained 2D Resnet 18 model are used to initialize the kernels, whereby each of the $t$ planes in the $t \times k \times k$ kernel are initialized with the pre-trained $k \times k$ weights, rescaled by $1/t$. This initialization setup produces the same results as the 2D pre-trained model run on a single static frame, repeated in the time domain.

In our setup, the first convolutional kernel in each Resnet layer is initialized with a $3 \times 3 \times 3$ kernel. The rest of the kernels are simply extended to their 3D counterpart that perform the same operation, by initializing them as a $1 \times 3 \times 3$ kernel. This approach does not increase the parameters significantly, thus alleviating overfitting.

### *2.4 Spatial Channel Image Fastformer (SHIFT) block*

In this section, we introduce our new Image/Video attention block called SHIFT (Spatial Channel Image Fastformer) and describe its operation. In self-attention, dot-product attention mechanism is used to fully model the interaction between query and key values through a similarity metric to determine which key regions are most similar to the query region. The complexity of this process requires an immense amount of resources, making it quite impractical for large inputs and especially 3D inputs. In our application, we have 3D sequences of screening mammograms, whereby we regard the time dimension (i.e., the screening sequence) as the third dimension in the video. Due to the increasing size of the input when we increase the number of sequential screening mammograms, the memory cost and computational complexity of the attention mechanism increases exponentially.

Observing the critical drawback of classical self-attention, this paper proposes an image/video based fast attention mechanism, which is inspired by the attention mechanism introduced in Fastformer [21] for text modelling, which uses additive attention. To the best of our knowledge, this is the first instance where additive attention model is investigated for images/videos, particularly in medical images. Additionally, in the text based Fastformer [21], the attention is only between different words. Our proposed Image Fastformer extends the additive attention to both spatial and channel attention mechanisms. From our previous studies [22, 23] and from the literature (e.g., Squeeze and Excitation network, Dual Attention Network [24, 25]), it can be observed that channel importance is also an important component in improving network performance.

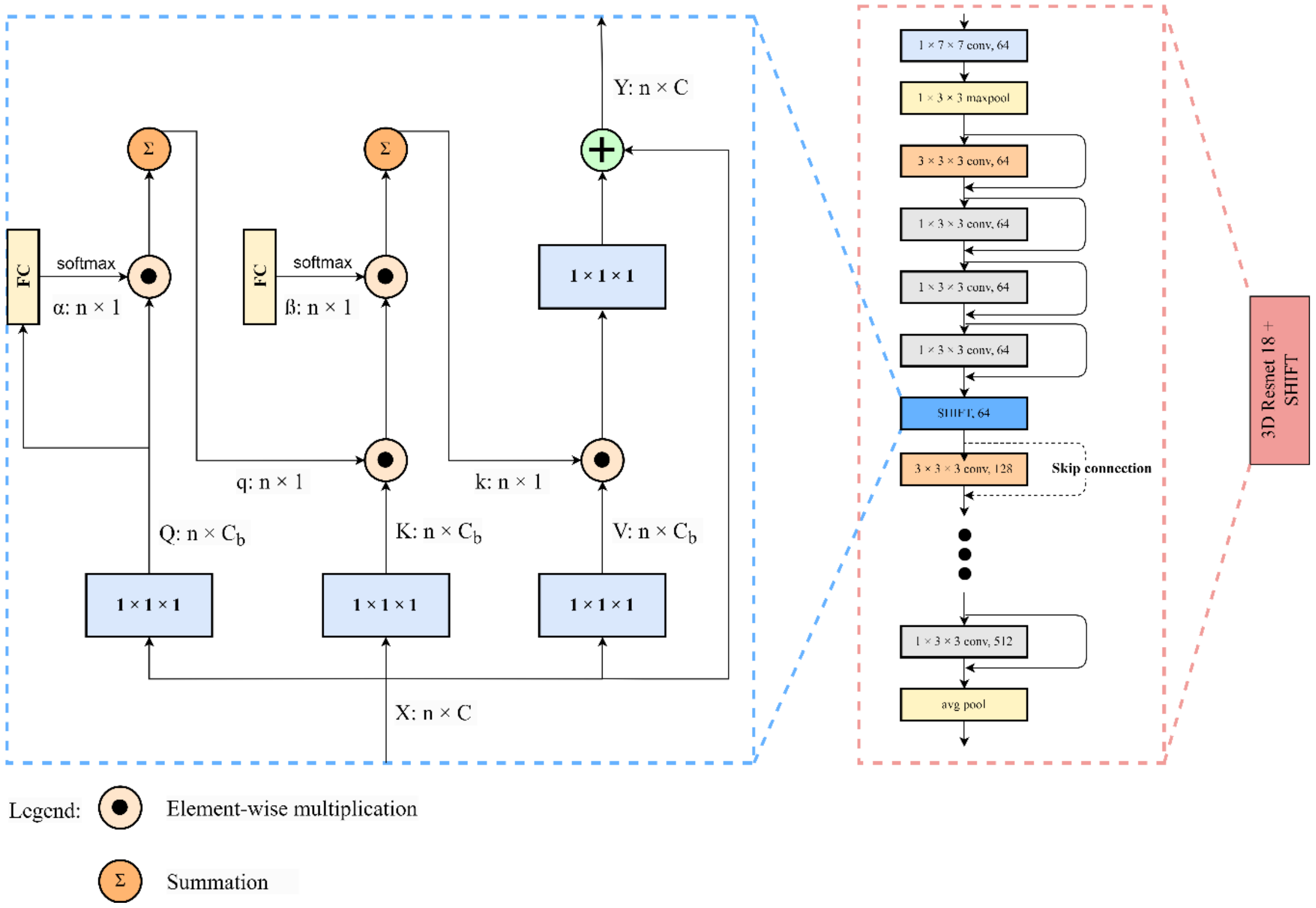

Fig. 3. Block diagram of our new *SHIFT* block and its placement in the 3D Resnet-18 network.

Similar to non-local networks [16], we utilize three distinct linear layers to convert the given input $X \in \mathbb{R}^{n \times C}$ into query, key, and value matrices $Q, K, V \in \mathbb{R}^{n \times C_b}$ where $n = T \times H \times W$ and $C_b$ denotes the internal (bottleneck) channel dimension. To summarize the encoded context in the query matrix, we first calculate the scalar values, $\alpha$ which are used to compute a weighted sum of the important regions:

$$\alpha = softmax\left(FC_Q(Q)\right) \in \mathbb{R}^{n \times 1} \tag{1}$$

where $FC_Q \in \mathbb{R}^{C_b \times 1}$ is a fully connected layer. The $\alpha$ values are the attention weights and will be responsible for identifying critical regions across the multiple (sequential) mammograms that the model should pay attention to. This step also circumvents the need to create the attention matrix of size $n \times n$ of exponential size/complexity. The global attention query is then calculated as:

$$q = \sum_{j=1}^{C_b} \alpha Q_j \in \mathbb{R}^{n \times 1}, \tag{2}$$

which summarizes the encoded context in the query. Summation along the channel dimension is intuitive, as opposed to summation along the voxel dimension $n$ as the latter loses critical information from the feature map. This is different from the original Fastformer model, which forms its global vectors through summation along the 1D (textual) sequence length.

We then pass this global information to the keys through element-wise product and model their interaction, thus allowing us to generate a global context-aware key matrix $p$:

$$p = q * K \in \mathbb{R}^{n \times C_b} \tag{3}$$

Similar to the query, we calculate an attention weight $ß$ using the global context-aware key matrix $p$:

$$ß = softmax\left(FC_K(p)\right) \in \mathbb{R}^{n \times 1}, \tag{4}$$

where $FC_K \in \mathbb{R}^{C_b \times 1}$ is a fully connected layer. This encapsulates the relevant query-key interactions which would then be passed to the value matrix. A crucial point to note is that both $FC_Q$ and $FC_K$ address the channel dimension efficiently by using each channel as attention values/weightage for all pixel positions. Combining these features through a weighted sum creates a global query/key vector.

Having two fully connected layers to weight the importance of each channel is also important as replacing $FC_K$ with $FC_Q$ depreciates the performance of the attention mechanism. We theorize that the two fully connected layers are responsible for summarizing different information from query and query-key interactions.

The global key can be calculated as such using the attention weight *ß*:

$$k = \sum_{j=1}^{C_b} ßK_j \in \mathbb{R}^{n \times 1} \quad (5)$$

Next, likewise with query-key interaction, we also model key-value interaction through element-wise multiplication of $k$ and $V$:

$$u = k * V \in \mathbb{R}^{n \times C_b} \quad (6)$$

This allows for value vectors to interact with global key and global query vector. Lastly, we pass the value matrix $u$ through a linear transformation layer to obtain the same channel dimension as the input. The attention features are then added back to the image features to form the final output. The query addition in the original Fastformer is omitted as it is redundant, as observed in our experiments. This block will be collectively known as the SHIFT (**S**patial C**h**annel **I**mage **F**as**t**former) block.

*2.5 Radiomics fusion for deep learning*

Radiomics is the process of extracting quantitative features from medical images, such as mammograms, using computer algorithms [26]. These features can provide additional information about the tissue structure and can be used for diagnosis, prognosis, and treatment planning. In mammography, radiomics features can be used to classify malignant and benign lesions [27, 28] and to predict breast cancer risk [15, 29].

Many studies have shown the potential of radiomics in supplementing traditional mammography by providing additional information about tissue structure and heterogeneity [4]. Combining deep learning features with radiomic features seems to be the natural progression to improve breast cancer risk prediction.

One key advantage of radiomics is that additional data collection is not needed as it can be generated from existing mammography data. This is more convenient when compared to collection of other risk factors such as age, reproductive history, genetic mutations, hormonal factors, family history, etc. Furthermore, these data are frequently missing, etc. and are not easily collated. Radiomics features are also easily extracted using the open-source Python library, Pyradiomics [30]. The majority of the features extracted in our study are obtained through the Pyradiomics [30] library and the usage of an open-source library will facilitate replicability of our described methods. Table 2 displays the feature groups/types from the Pyradiomics toolbox and their corresponding numbers and descriptions that we utilized for risk prediction.

Table 2. Computed radiomics features from the Pyradiomics library according to grouping or type, and their description.

| Feature group/type | Feature number/quantity | Description |
|---|---|---|
| First Order | 18 | Energy, Total Energy, Entropy, Minimum, 10$^{th}$ Percentile, 90$^{th}$ Percentile, Maximum, Mean, Median, Interquartile Range, Range, Mean Absolute Deviation, |
| | | Robust Mean Absolute Deviation, Root Mean Squared, Skewness, Kurtosis, Variance, Uniformity |
| Gray Level Co-occurrence Matrix (**GLCM**) | 23 | Autocorrelation, Joint Average, Cluster Prominence, Cluster Shade, Cluster Tendency, Contrast, Correlation, Difference Average, Difference Entropy, Difference Variance, Joint Energy, Joint Entropy, Informational Measure of Correlation 1, Informational Measure of Correlation 2, Inverse Difference Moment, Maximal Correlation Coefficient, Inverse Difference Moment Normalized, Inverse Difference, Inverse Difference Normalized, Inverse Variance, Maximum Probability, Sum Entropy, Sum Squares |
| Gray Level Size Zone Matrix (**GLSZM**) | 16 | Small Area Emphasis, Large Area Emphasis, Gray Level Non-Uniformity, Gray Level Non-Uniformity Normalized, Size Zone Non-Uniformity, Size Zone Non-Uniformity Normalized, Gray Level Variance, Zone Variance, Zone Entropy, Low Gray Level Zone Emphasis, High Gray Level Zone Emphasis, Small Area Low Gray Level Emphasis, Small Area High Gray Level Emphasis, Large Area Low Gray Level Emphasis, Large Area High Gray Level Emphasis |
| Gray Level Run Length Matrix (**GLRLM**) | 14 | Short Run Emphasis, Long Run Emphasis, Gray Level Non-Uniformity, Gray Level Non-Uniformity Normalized, Run Length Non-Uniformity, Run Length Non-Uniformity Normalized, Run Percentage, Gray Level Variance, Run Variance, Run Entropy, Low Gray Level Run Emphasis, High Gray Level Run Emphasis, Short Run Low Gray Level Emphasis, Short Run High Gray Level Emphasis, Long Run Low Gray Level Emphasis, Long Run High Gray Level Emphasis |
| Neighbouring Gray Tone Difference Matrix (**NGTDM**) | 5 | Coarseness, Contrast, Busyness, Complexity, Strength |
| Gray Level Dependence Matrix (**GLDM**) | 15 | Small Dependence Emphasis, Large Dependence Emphasis, Gray Level Non-Uniformity, Dependence Non-Uniformity,<br>Dependence Non-Uniformity Normalized, Gray Level Variance,<br>Dependence Variance, Dependence Entropy, Low Gray Level Emphasis, High Gray Level Emphasis, Small Dependence Low Gray Level Emphasis, Small Dependence High Gray Level Emphasis, Large Dependence Low Gray Level Emphasis,<br>Large Dependence High Gray Level Emphasis |

Besides the standard radiomic features, we added new features to improve our model's capability: Supplemented features include statistical features calculated from two image maps in the frequency domain using Discrete Cosine Transform (DCT) and Fast Fourier transform (FFT). Adoption of these features have shown to improve breast cancer benign/malignant classification [27] and should also enhance risk assessment.

Let be $f(x, y)$ be the input image with size $M$ by $N$, 2D DCT can be formulated as:

$$DCT(m,n) = \frac{2}{\sqrt{MN}} C(m)C(n) \sum_{x=0}^{M-1} \sum_{y=0}^{N-1} f(x,y) \cos(\frac{(2x+1)m\pi}{2M}) \cos(\frac{(2y+1)n\pi}{2N}), \tag{7}$$

where $C(m) = C(n) = 1/\sqrt{2}$ for $m,n = 0$ and $C(m), C(n) = 1$ otherwise. Then, 2D FFT can be formulated as:

$$F(m,n) = \sum_{x=0}^{M-1} \sum_{y=0}^{N-1} f(x,y) \exp(-j2\pi(\frac{xm}{M} + \frac{yn}{N})), \tag{8}$$

where $0 \leq m \leq M-1$ and $0 \leq n \leq N-1$.

As key information can be extrapolated from the distribution patterns of fatty and fibro-glandular tissues in the breast, further analysis of parenchymal texture in a mammogram is important. Specific patterns can be revealed in the local spectral or frequency content of the mammogram. Most information about the mammogram is usually centred around the low frequency components, while noise tend to be found in the higher frequencies. From the two computed DCT and FFT frequency domain image maps, we computed the statistical features listed in Table 3.

Table 3. New DCT and FFT frequency domain statistical features computed for cancer risk prediction.

| Feature group/type | Feature number/quantity | Description |
|---|---|---|
| Discrete Cosine Transform (**DCT**)<br>Fast Fourier Transform (**FFT**) | 30 | Mean, Maximum, Variance, Skew, Kurtosis, Entropy, Energy, Root Mean Square, Uniformity, Minimum, Median, Range, Interquartile Range, Mean Absolute Deviation, Median Absolute Deviation |

Previous studies have shown that the addition of clinical risk factors, e.g., woman's age, family history, subjective breast density BIRADS rating can help improve the risk model’s performance [2, 29]. Thus, in our study, we included patient’s age in our model; we did not have access to other clinical factors due to ethical/privacy restrictions.

We included all computed radiomics features alongside patient’s age in the fully connected layer of our model. We had initially experimented various approaches in the literature to incorporate the radiomics and clinical features including Sequential Floating Forward Selection (SFFS), Sequential Floating Backward Selection (SFBS), Dynamic Affine Feature Map Transform (DAFT) [31], etc. However, our experiments showed that just including all features into the fully connected layer produced the best result. This lightweight approach is effective, and the weights of relevant/useful features can be easily tuned in the fully connected layer with a low number of additional parameters that require training.

### *2.6 Multi-View Gating Mechanism*

In the literature, different methods and approaches have been proposed to combine image features from the four different mammographic views [32-34]. Almost all proposed approaches are quite complex and require fully connected layers that involve considerable parameter tuning and/or are computationally intensive. In this study, in line with our SHIFT method that is efficient with reduced parameters, we examine several lightweight but effective approaches to combine the different views.

In our initial approach, we utilize the simplest and most lightweight approach of averaging 4 outputs from the fully connected layer for each view. Then, the model’s output is calculated as follows:

$$y = \sigma\left(\frac{X_{LCC} + X_{RCC} + X_{LMLO} + X_{RMLO}}{4}\right), \quad (9)$$

where $\sigma$ denotes the sigmoid function while $X_{LCC}, X_{RCC}, X_{LMLO}, X_{RMLO}$ represent the output from the fully connected layer corresponding to each view and $y$ is the output of the model. This approach however gives equal importance to both CC and MLO views. Conversely, most results from literature including our previous studies show that certain views provide more information than others in predicting/detecting breast cancer [9, 32, 35].

To reweight the importance of each view, we add a gate to facilitate the model in deciding which view contributes more to risk assessment. This allows the model to stochastically determine whether to restrict or enhance the information from each of the views. The calculation of the model's output would be updated as such:

$$y = \sigma\left(\frac{W_T X_{LCC} + W_T X_{RCC} + W_S X_{LMLO} + W_S X_{RMLO}}{2W_T + 2W_S}\right), \quad (10)$$

$$W_T, W_S = W_f + \tanh(W_\theta), \quad (11)$$

where $W_T$ and $W_S$ represent the scale/gate for the CC and MLO views, respectively. In both the scales, $W_f$ is a fixed weight while $W_\theta$ is the trainable parameter. The fixed weight $W_f$ is there to prevent either scale from collapsing to the value of 0 during training, thus completely losing information from that view. Both scales are initially initialized to be equal to 1 to simulate the averaging setup of 4 views; refer to (9).

*2.7 Bilateral Asymmetry based Finetuning*

Bilateral asymmetry between breasts is often indicative of cancer development as cancer typically develop in only one breast [36]. Our previous studies [15, 29, 32] verified this and showed that risk scores computed by bilateral mammographic feature asymmetry based models have potential to predict short-term risk of women for developing breast cancer. Thus, we first remove the gate (in Section 2.6) temporarily and generate risk scores for each independent mammographic view. We then use these scores to calculate the difference of risk scores between the contralateral breasts, as follows:

$$y_{LCC} = \sigma(X_{LCC}), \quad y_{RCC} = \sigma(X_{RCC}), \quad y_{LMLO} = \sigma(X_{LMLO}), \quad y_{RMLO} = \sigma(X_{RMLO}), \quad (12)$$

$$\gamma = |(y_{LCC} + y_{LMLO}) - (y_{RCC} + y_{RMLO})|, \quad (13)$$

where $y_{LCC}, y_{RCC}, y_{LMLO}, y_{RMLO}$ represent risk scores of each left and right breast CC and MLO views from one patient, and $\gamma$ represents the asymmetry risk score between left and right breasts. The new asymmetry risk score, $\gamma$ indicates which patient has signs of bilateral asymmetry in their mammography image.

We first train our model using this method on the training subset. We then adopt a pseudo-labelling method, whereby we use our trained model to make predictions and generate soft labels on all of the images in the training set along with their corresponding asymmetry risk score. We retrain our classifier to fit the soft pseudo-labels obtained on the training set. In our experiments, we observe that the asymmetry risk score is on average higher in cases in the training set compared to controls, which confirms (with our previous studies) that bilateral asymmetry is higher in cases (due to cancers typically developing in only one breast) compared to controls.

In the context of our approach, we can formulate our sequential screening CSAW dataset as follows:

$$D_s = \{(x_i, y_i, \bar{y}_i, \gamma_i)\}_{i=1}^{n}, \quad (14)$$

where $(x_i, y_i, \bar{y}_i, \gamma_i)$ represents the image, true label, soft label, and asymmetry score of the $i^{th}$ patient, respectively in the dataset; $n$ represents the number of patients in the dataset.

The CSAW screening dataset is very imbalanced with 873 cases and 7850 controls, altogether, which can affect the training/learning ability of our model. Due to the dataset imbalance, we sought to reduce the number of controls in our dataset. Namely, we removed patients with a high asymmetry risk score, $\gamma$ by determining if the score exceeds a certain percentile $P_T$ within the controls of the training set, with the threshold percentage $T$ being a tunable hyperparameter. This eliminates any controls that are deemed inconsistent (i.e., highly asymmetrical/False Positive) by the model and allows us to form a filtered training set. which allows us to include bilateral asymmetry awareness to create a more ideal dataset for better finetuning. Cases are maintained in the dataset as their proportion is already so low and lowering the number of cases further would hinder the model's training performance. Thus, we obtain the filtered training dataset as follows:

$$D_f \subset D_s = \{(x_i, y_i, \bar{y}_i, \gamma_i) \mid \big((\gamma_i < P_T) \cap (y_i = 0)\big) \cup (y_i = 1)\}_{i=1}^{m}, \qquad (15)$$

where $m$ represents the total number of patients in the filtered dataset and $m < n$.

From experiments, we found that training entirely on either dataset $D_s$ or $D_f$ produces poor results. Training solely on the filtered dataset results in poor performance of the model due to a lower amount of training samples, while only training on the original dataset does not bode well due to the inclusion of controls with high asymmetrical risk scores. To remedy this, we instead pretrain our model on dataset $D_s$ first, for half the number of total epochs, then finetune our model on dataset $D_f$. This allows our model to take advantage of both datasets, to improve its performance. The entire process in visualized in Fig. 4.

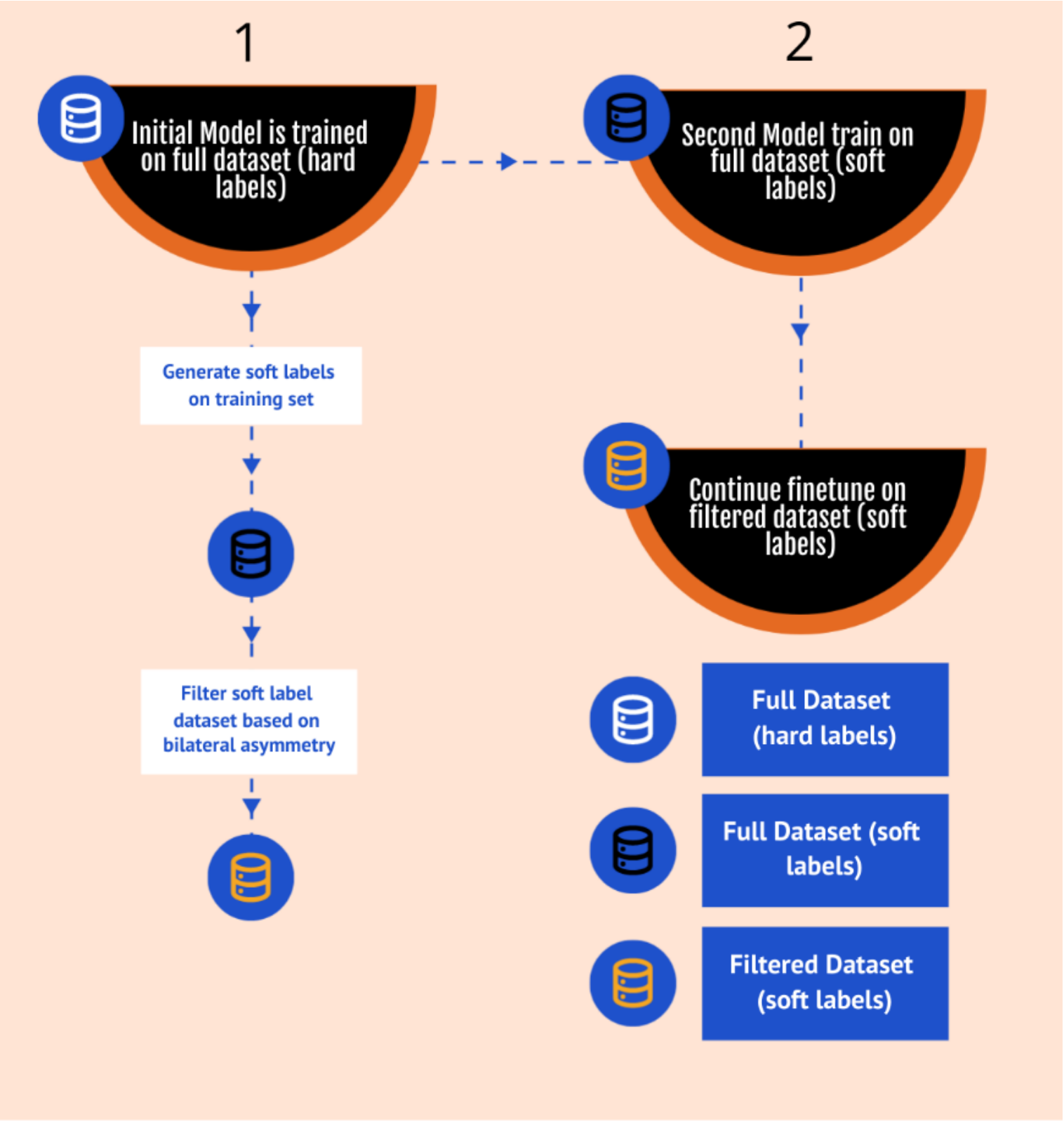

Fig. 4. Flowchart of bilateral asymmetry based finetuning.

*2.8 Proposed Architecture*

The block diagram of our proposed deep learning architecture is displayed in **Error! Reference source not found.** 5. We trained our network for 60 epochs with a batch size of 12 (number of patients) using the binary cross entropy loss. The Adam optimizer [37] was used to train our model with a learning rate of 0.0001. The training was performed on two NVIDIA RTX 2080 Ti GPUs.

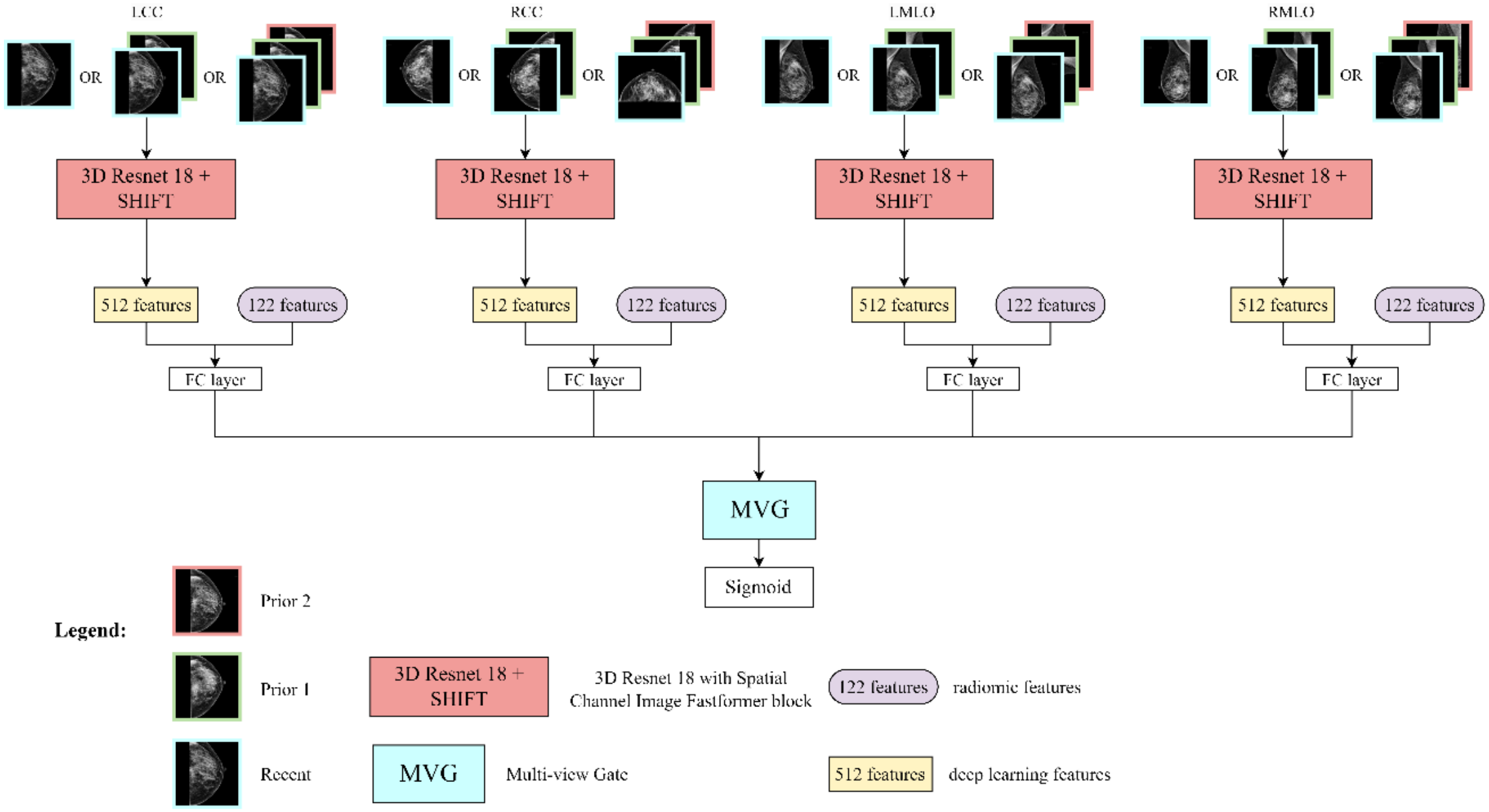

Fig. 5. The proposed deep learning architecture that can capture the changes in breast tissue structure over time through SHIFT block for predicting breast cancer risk, with the number of varying screenings (1 to 3). The model takes in mammograms from 4 views along with their Radiomics features (122 new and established features) in the fully connected layer. A new gating mechanism and a new bilateral asymmetry based finetuning approach are implemented to output the final risk score. The weights of the 3D Resnet 18 + SHIFT and fully connected layer are shared across all 4 views.

### *2.9 Experimental Setup and Classification Methodology*

We randomly obtained a 20% split of our dataset to form an independent test set consisting of 19,652 mammograms from 1,749 women. A patient-wise stratified 5-fold cross-validation (CV) was performed on the rest of the dataset consisting of 79,136 mammograms from 6,974 women. In each CV run, the network parameters were optimized on the training set (4 folds) and the remaining fold constituted the validation set. The proportions of cases and controls were maintained throughout the training, validation, and testing sets as much as possible, to maintain consistency in the dataset splits.

Following the same evaluation method in [8], to compute the 2-year AUC results, cases are considered as positive if their prior year mammograms are followed by a cancer diagnosis within 2 years. Similar evaluation methods were used to compute the 1-year AUC and > 2-year AUC results. Thus, we evaluate the anonymized ground truth results as 1-year AUC, 2-year AUC and > 2-year AUC, corresponding to the provided ground truth labels in the dataset. All the results are tabulated in corresponding tables in the Results section. We utilized bootstrapping methods to obtain the 95% confidence intervals (CI) for all the AUC values. All scripts and our CNN architecture are implemented using Python package (v3.8.5), Scikit-learn (v1.0.2), Matplotlib (v3.5.1) and PyTorch (v1.10.1) libraries.

## 3. RESULTS

Table 4 tabulates and compares the results of different numbers of mammographic screenings on the baseline 3D Resnet-18 model. The model used here is the baseline Resnet-18 architecture with patient risk scores averaged from the 4 views. From the results, we can observe that providing additional screening mammograms to the model enables it to better predict cancer risk. Namely, as the number of screenings increase from 1 to 3, the AUC gradually increases. This indicates that the 3D convolutional kernels can capture spatiotemporal relationships between the sequential mammograms, thus producing overall better results.

Table 4. Tabulated results for 3 different categories of patients diagnosed with cancer in the CSAW dataset corresponding to 1-year AUC, 2-year AUC and > 2-year AUC, on the baseline three-dimensional (3D) Resnet-18 network with 95% confidence intervals (CI). The table also tabulates the results obtained by taking different numbers of mammographic screenings (i.e., 1, 2 or 3 screenings), from the dataset.

| Model | Number of Screenings | 1-year AUC (95% CI) | 2-year AUC (95% CI) | > 2-year AUC (95% CI) |
|---|---|---|---|---|
| Baseline | 3 | 0.882 (0.853-0.913) | 0.855 (0.829-0.886) | 0.849 (0.823-0.878) |
| | 2 | 0.879 (0.85-0.91) | 0.839 (0.811-0.871) | 0.835 (0.81-0.866) |
| | 1 | 0.877 (0.849- 0.909) | 0.831 (0.798-0.863) | 0.823 (0.793-0.853) |

For the rest of the experiments however, we utilized only 2 mammographic screenings to form the input. Even though using 3 mammogram screenings produces good results in Table 4, we risk increasing the amount of duplicated mammograms in the input video, as not all patients went for 2 or more sequential screenings (as depicted in Fig. 1). As explained in Section 2.2 (Data Preprocessing), for patients with insufficient screenings, we duplicated their most recent screenings to fill the timepoint gaps in the video form of the sequential mammograms. Due to the low numbers of 3 or more sequential screenings in our dataset (see Fig. 1), we used only 2 mammographic screenings, to avoid duplicating the mammograms to maintain the accuracy and performance of our models.

Table 5 presents a comparison of results obtained from various configurations of our SHIFT block in layer 1 of the model. We investigated the use of weights sharing for the query/key convolutions and the fully connected layers of alpha/beta in the SHIFT block. Upon analysing the results, it becomes evident that weights sharing, while reducing the number of trainable weights, diminishes the performance of our attention block. This suggests the importance of separate trainable weights for query/key and alpha/beta, as they are responsible for capturing more information from the image features. Additionally, we experimented with query and value addition, as described in the original paper [21]. While the outcomes were promising, our final configuration (illustrated in Fig. 3) remains the optimal setup.

Table 5. Summary of results for SHIFT in layer 1 with different configurations

| Configuration | 1-year AUC (95% CI) | 2-year AUC (95% CI) | > 2-year AUC (95% CI) |
|---|---|---|---|
| SHIFT | **0.887** (0.86-0.917) | **0.848** (0.82-0.876) | **0.844** (0.818-0.872) |
| Weight sharing – Query, Key | 0.881 (0.852-0.914) | 0.835 (0.803-0.865) | 0.831 (0.802-0.86) |
| Weight sharing – Alpha, Beta | 0.873 (0.844-0.903) | 0.841 (0.815-0.875) | 0.838 (0.808-0.866) |
| Query Value addition | 0.879 (0.848-0.911) | 0.841 (0.81-0.874) | 0.84 (0.813-0.867) |

Table 6 tabulates the performance of the different attention mechanisms of Non-Local Networks and our SHIFT block in our deep learning architecture. Our results show that adding an attention mechanism improves the performance of our risk prediction models; this is true for both Non-Local Networks and SHIFT, whereby AUC results improved for all 1-year, 2-year and > 2-year AUC.

We also investigated the inclusion of the attention block at different layers of Resnet in Table 6. To obtain the highest effect/benefit from the attention mechanism, we hypothesised that higher-resolution image features would benefit more from the attention mechanism, as applying the attention early on will subsequently pervade/benefit the lower-resolution image features. The results validated our hypothesis, whereby the attention blocks placed at layer 1 of Resnet outperformed layer 2 for SHIFT (for Non-Local network, there was insufficient memory when the block was placed in layer 1, due to the exponential increase in computations, especially for higher screening mammogram numbers). Our SHIFT method outperforms Non-Local attention for all tabulated AUC results.

Table 6. Summary of results for Non-Local networks, SHIFT and no attention mechanisms applied. For Non-Local and SHIFT, the corresponding attention blocks were added into 2 different stages, i.e., layer 1 or layer 2 of Resnet-18.

| Attention Mechanism | Resnet Layer | 1-year AUC (95% CI) | 2-year AUC (95% CI) | > 2-year AUC (95% CI) |
|---|---|---|---|---|
| - | - | 0.879 (0.85-0.91) | 0.839 (0.811-0.871) | 0.835 (0.81-0.866) |
| Non-Local | 1 | | Out of Memory | |
| | 2 | 0.881 (0.822-0.91) | 0.840 (0.811-0.875) | 0.837 (0.81-0.867) |
| SHIFT | 1 | **0.887** (0.86-0.917) | **0.848** (0.82-0.876) | **0.844** (0.818-0.872) |
| | 2 | 0.880 (0.85-0.911) | 0.842 (0.811-0.873) | 0.839 (0.812-0.867) |

For an input of the same dimension, SHIFT is significantly more efficient in terms of computation as compared to Non-Local. This is due to the advantage of SHIFT utilizing additive attention to summarize image information. SHIFT avoids computing the similarity matrix between each pair of image pixel positions, which greatly reduces both memory and computational costs. To examine this, we calculate the computational costs of both Non-Local and SHIFT when the attention block is inserted in the same Resnet layer (layer 1), as tabulated in Table 7. At the cost of a slight increase in the number of parameters (in the fully connected layers), SHIFT is significantly more efficient and faster.

Table 7. Comparison of resource usages of Non-Local and SHIFT modules in our three-dimensional (3D) network architecture.

| | Non-Local | SHIFT |
|---|---|---|
| Computation (MAC): | 268.427M | **67.682M** |
| Number of parameters: | 8.192K | 8.258K |

MAC = Multiply-Add cumulation

Fig. 6 illustrates two attention maps computed by the non-local block and our SHIFT block for a cancer case in our dataset. These attention maps visually depict which part of the current (left) and prior 1 (right) image the attention mechanism pays attention to. For Non-local, a query position around the dense tissue region is used and 10 highest weighted points are visualized. On the other hand, SHIFT utilizes attention with a dimension of $\mathbb{R}^{n \times 1}$. To visualize this, 20 largest weighted points on the attention map are plotted instead, allowing us to grasp the key areas of focus for the SHIFT attention mechanism. It can be observed that the SHIFT attention mechanism focuses on the dense regions in the current and prior images, whereas Non-local attends to the fatty and boundary regions of the images.

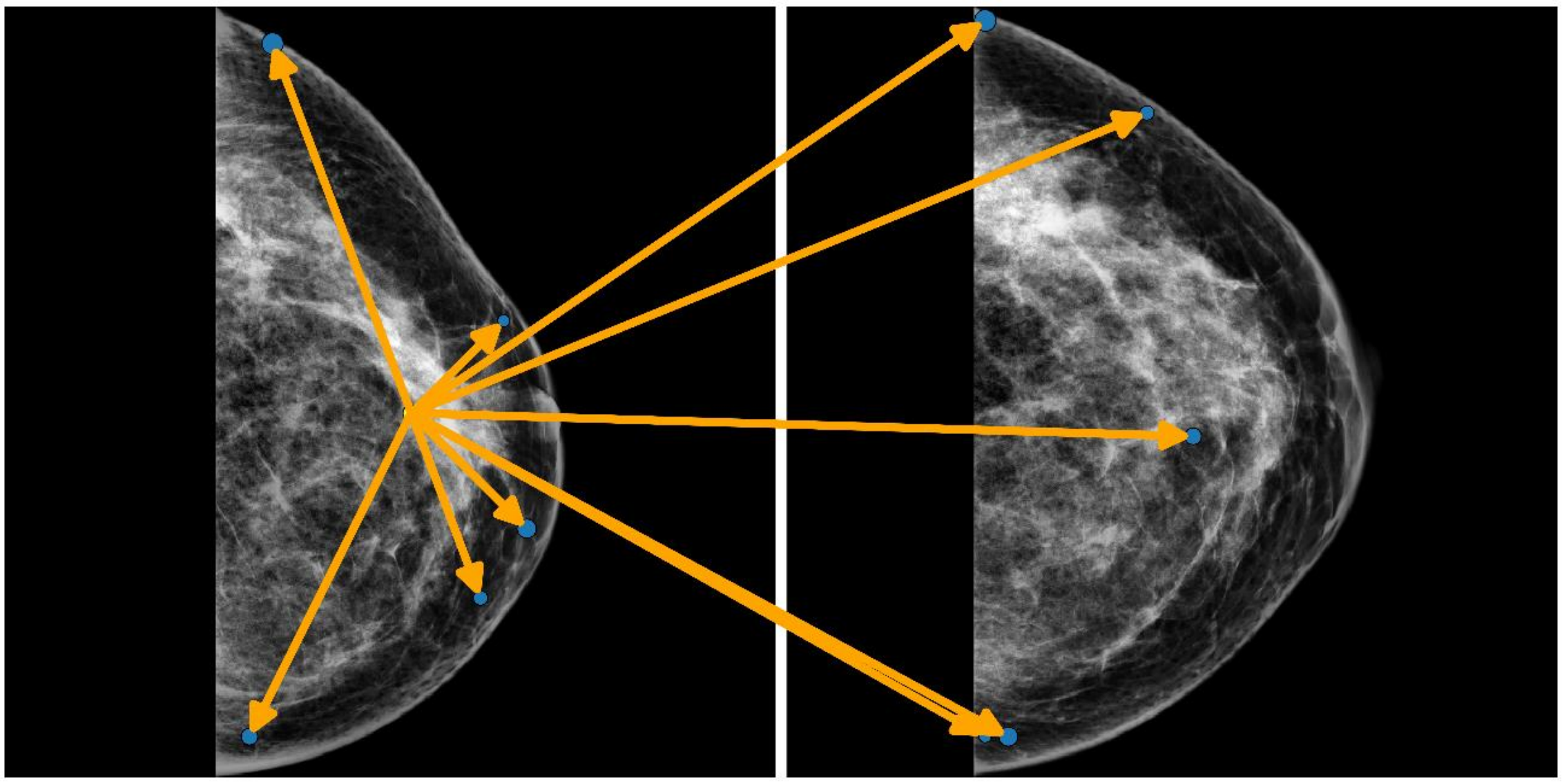

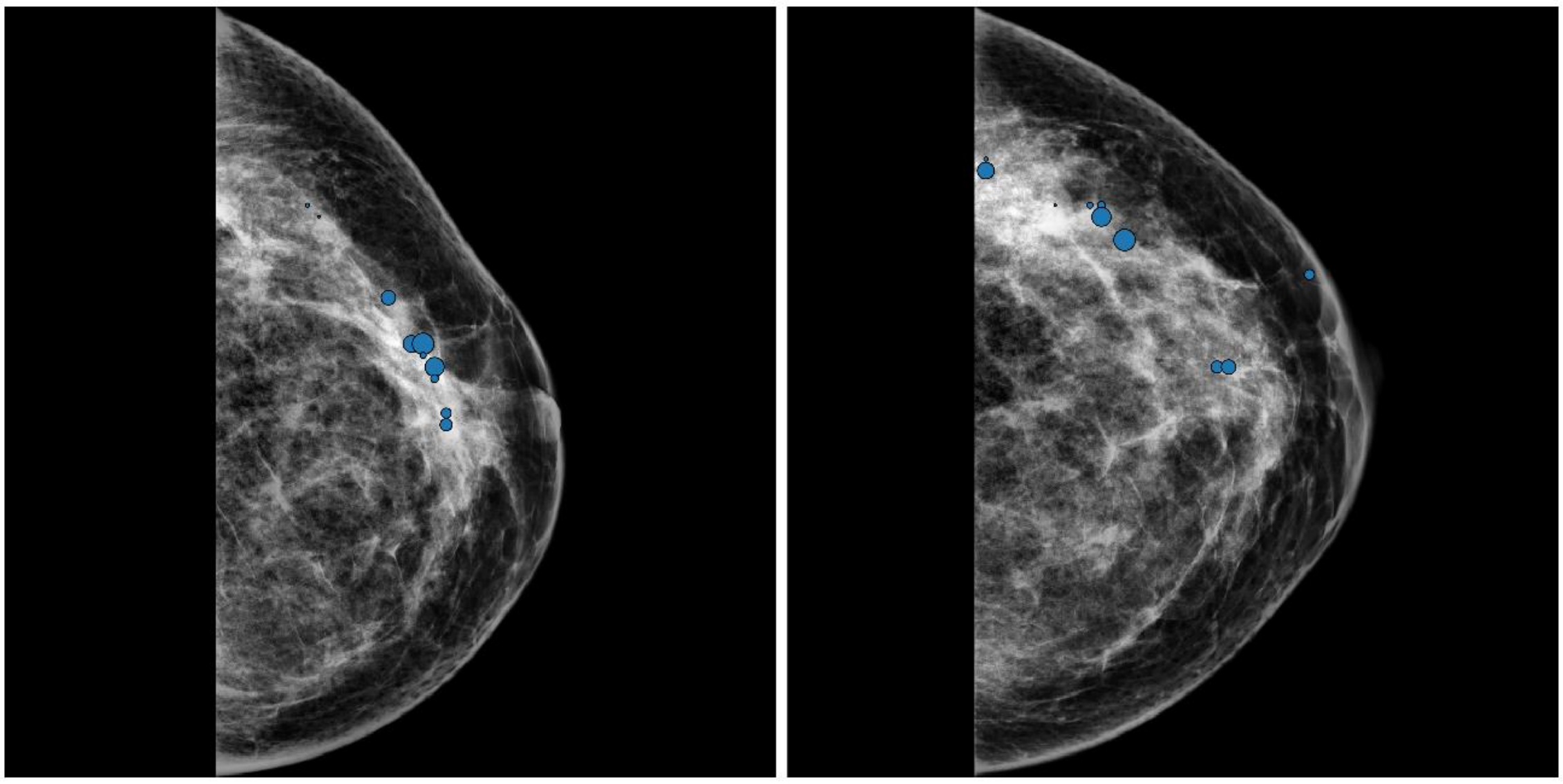

Fig. 6. (Top) Attention map of Non-local attention. The arrows indicate the flow of attention from a starting position (queried position) to the ending points of the current (left) and prior 1 (right) images of a cancer patient in our dataset. The 10 highest weighted points are visualized here. (Bottom) Attention map alpha of SHIFT on the same patient. Weighted points go from largest to smallest in descending order and across both mammogram inputs.

Table 8 tabulates the performance of radiomic features when included in our model. We included radiomic features by concatenating them with the deep learning features before passing them through the fully connected layer. Our test shows that the inclusion of our frequency based features improves our model's performance. This is particularly valuable in situations where gathering a collection of clinical risk factors may not be practical or feasible. In our specific case, leveraging radiomic features proves to be instrumental in bolstering the predictive power of our model.

Table 8. Summary of results for the addition of radiomic features into our model. The radiomic features are concatenated with the deep learning features in the fully connected layer.

| Radiomic Features | | 1-year AUC (95% CI) | 2-year AUC (95% CI) | > 2-year AUC (95% CI) |
|---|---|---|---|---|
| Pyradiomics | Frequency | | | |
| - | - | 0.879 (0.85-0.91) | 0.839 (0.811-0.871) | 0.835 (0.81-0.866) |
| ✓ | - | 0.875 (0.842-0.91) | 0.841 (0.811-0.877) | 0.841 (0.814-0.87) |
| ✓ | ✓ | **0.880** (0.852-0.909) | **0.845** (0.817-0.876) | **0.844** (0.818-0.87) |

In addition to our previous findings, we explored an alternative feature combination approach called DAFT [31], which we implemented into the third layer of our Resnet architecture. To evaluate its effectiveness, we compared the performance of our model using different methods of feature combination along with SHIFT, as outlined in Table 9. Although our results indicated that DAFT is a viable method, we ultimately decided to proceed with the initial feature combination approach, as it yielded the best overall performance.

Table 9. Summary of results for different combination methods of radiomic features with the addition of SHIFT.

| Combination method | SHIFT | 1-year AUC (95% CI) | 2-year AUC (95% CI) | > 2-year AUC (95% CI) |
|---|---|---|---|---|
| - | - | 0.879 (0.85-0.91) | 0.839 (0.811-0.871) | 0.835 (0.81-0.866) |
| Fully Connected layer | - | 0.880 (0.852-0.909) | 0.845 (0.817-0.876) | 0.844 (0.818-0.87) |
| | ✓ | **0.888** (0.86-0.919) | 0.854 (0.825-0.882) | **0.853** (0.827-0.877) |
| DAFT [31] | - | 0.878 (0.85-0.914) | **0.858** (0.832-0.888) | 0.852 (0.829-0.879) |
| | ✓ | 0.886 (0.856-0.914) | 0.857 (0.829-0.889) | 0.853 (0.829-0.878) |

Moving on, Table 10 presents the addition of gates into our model which enables us to effectively weigh information from both views. To evaluate the impact, we experimented with different starting fixed/trainable weights to observe its effect. Our results show that the addition of a gate improves the performance of our risk prediction models irrespective of the chosen starting weights with some performing better. Interestingly, the results consistently reveal a notable preference for the CC view over the MLO view within our model, which is consistent with previous studies [9, 32], possibly due to the presence of pectoral muscle in the MLO view which may confound the model.

Table 10. Comparison of different starting weights for the gating mechanism.

| Gate Weights | | 1-year AUC (95% CI) | 2-year AUC (95% CI) | > 2-year AUC (95% CI) |
|---|---|---|---|---|
| $W_T, W_S$ | $W_f$ | | | |
| 0.5 | 0.5 | 0.888 (0.858-0.918) | 0.86 (0.833-0.891) | 0.858 (0.832-0.886) |
| 0.6 | 0.4 | **0.902** (0.877-0.928) | **0.863** (0.835-0.891) | **0.861** (0.833-0.886) |
| 0.7 | 0.3 | 0.894 (0.868-0.925) | 0.86 (0.83-0.889) | 0.855 (0.83-0.883) |
| 0.8 | 0.2 | 0.886 (0.856-0.916) | 0.863 (0.837-0.892) | 0.859 (0.832-0.885) |

$W_T$ = CC weight, $W_S$ = MLO weight, $W_f$ = fixed weight

Utilizing our best model, we begin by reconstructing the complete dataset using soft labels. Next, we apply the bilateral risk score technique to identify and remove outlier controls, resulting in a refined filtered dataset. As presented in Table 11, demonstrates that training exclusively on either of these datasets yields unsatisfactory model performance. We can observe some improvement in the 2-year AUC and >2-year AUC when the model is trained on the full dataset, which could be attributed to the soft labels. However, the optimal approach involves leveraging both datasets during training, leading to superior model performance, as evidenced by the results in Table 11.

Table 11. Performance of our model on the different versions of dataset. The final iteration was first trained on the full dataset, then finetuned on the filtered dataset.

| Dataset | 1-year AUC (95% CI) | 2-year AUC (95% CI) | > 2-year AUC (95% CI) |
|---|---|---|---|
| $D_s$ | 0.898 (0.871-0.929) | 0.867 (0.839-0.897) | 0.862 (0.837-0.888) |
| $D_f$ | 0.898 (0.871-0.925) | 0.86 (0.832-0.889) | 0.855 (0.829-0.882) |
| $D_s \rightarrow D_f$ | **0.905** (0.88-0.931) | **0.872** (0.846-0.898) | **0.866** (0.844-0.89) |

$D_s$ = Full dataset, $D_f$ = Filtered dataset

Table 12 tabulates the results of incorporating new modifications to our model. To compare our method to the state-of-the-art Mirai method for breast cancer risk prediction, we replicated the authors' setup in [8] as closely as possible using the descriptions and hyperparameters provided in their paper [8] and their codes provided on GitHub, and trained their model on our training set for a fair comparison. Our final model's performance is better, which could be mainly attributed to effectively utilizing the patient's imaging history (namely, the previous image screening(s)) in our model. This was also acknowledged by the authors as a future direction/approach to improve Mirai's performance [8]. To statistically validate our results, we conducted DeLong's test [38], utilizing our model as the reference and comparing it against Mirai. The test yielded $p = 0.03$, $p < 0.01$ and $p < 0.01$ for 1-year, 2-year and > 2-year AUC respectively, indicating a significant difference between the two models.

The results show that the final model that combines deep learning, attention mechanism, radiomic features, gating and bilateral finetuning produces the best AUC results. The results also show that the self-attention mechanism of the SHIFT blocks can capture useful information across different sequential screening mammograms, to predict cancer risk in the most recent screening mammogram, and both deep learning and radiomic features can work hand-in-hand to provide complementary/supplementary information for cancer risk prediction. The addition of gates to reweight the contribution of each mammographic view is beneficial to the model. Our results also show that assimilating knowledge of bilateral asymmetry to finetune our model can further improve performance. Receiver Operating Characteristic (ROC) curves for 1-year, 2-year and >2-year AUC results are plotted in Fig. 3.

Table 12. Compilation of results of incorporating new modifications/additions to our model.

| Model Configuration | Radiomics | 1-year AUC (95% CI) | 2-year AUC (95% CI) | > 2-year AUC (95% CI) |
|---|---|---|---|---|
| SHIFT + Gate + BAF | ✓ | **0.905** (0.88-0.931) | **0.872** (0.846-0.898) | **0.866** (0.844-0.89) |
| SHIFT + Gate | ✓ | 0.902 (0.877-0.928) | 0.863 (0.835-0.891) | 0.861 (0.833-0.886) |
| SHIFT | ✓ | 0.888 (0.86-0.919) | 0.854 (0.825-0.882) | 0.853 (0.827-0.877) |
| Baseline | ✓ | 0.880 (0.852-0.909) | 0.845 (0.817-0.876) | 0.844 (0.818-0.87) |
| Baseline | - | 0.879 (0.85-0.91) | 0.839 (0.811-0.871) | 0.835 (0.81-0.866) |
| Mirai [8] | - | 0.877 (0.848-0.911) | 0.834 (0.804-0.865) | 0.830 (0.802-0.859) |

SHIFT = Spatial Channel Image Fastformer, BAF = Bilateral Asymmetry based Finetuning

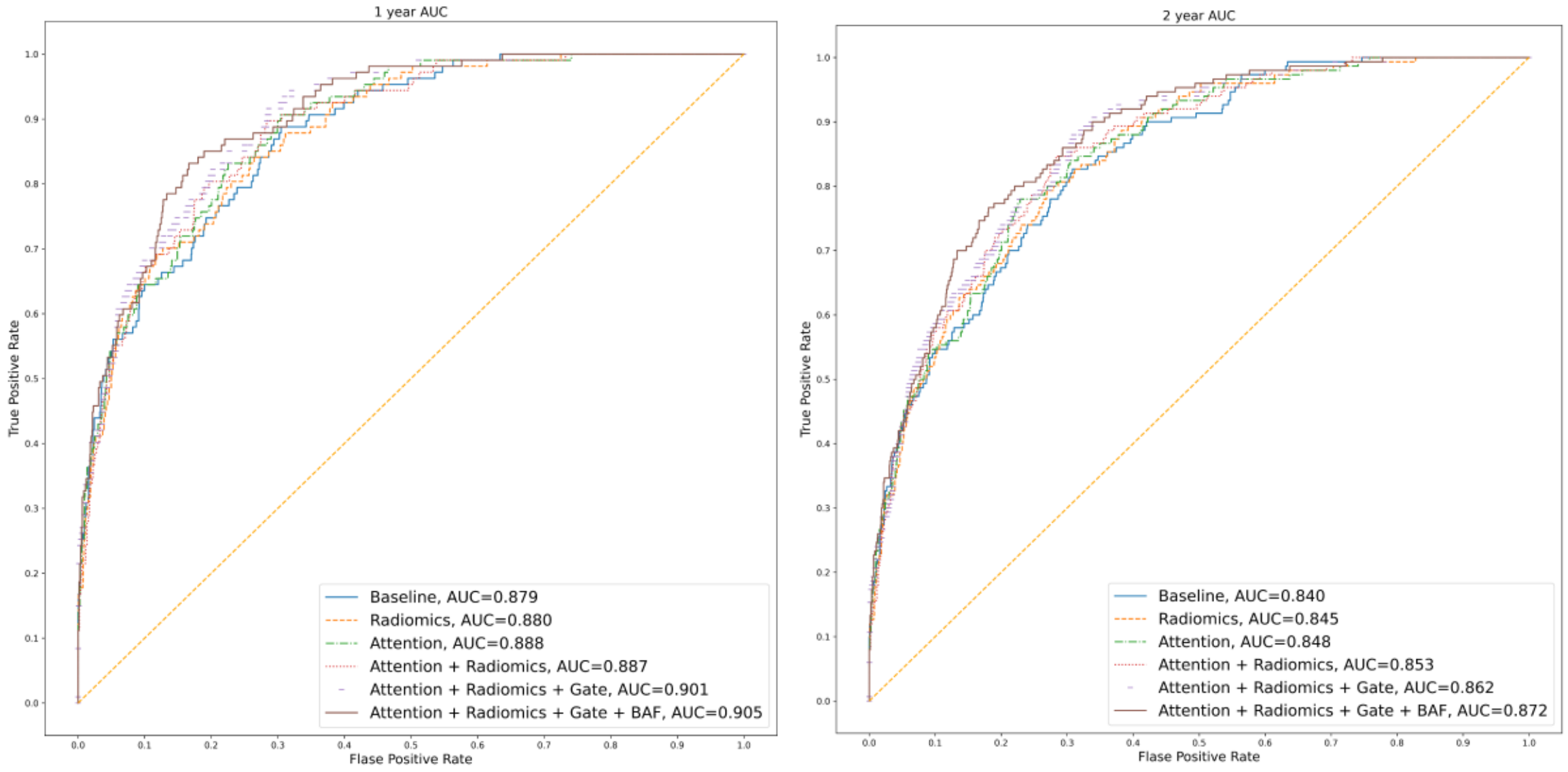

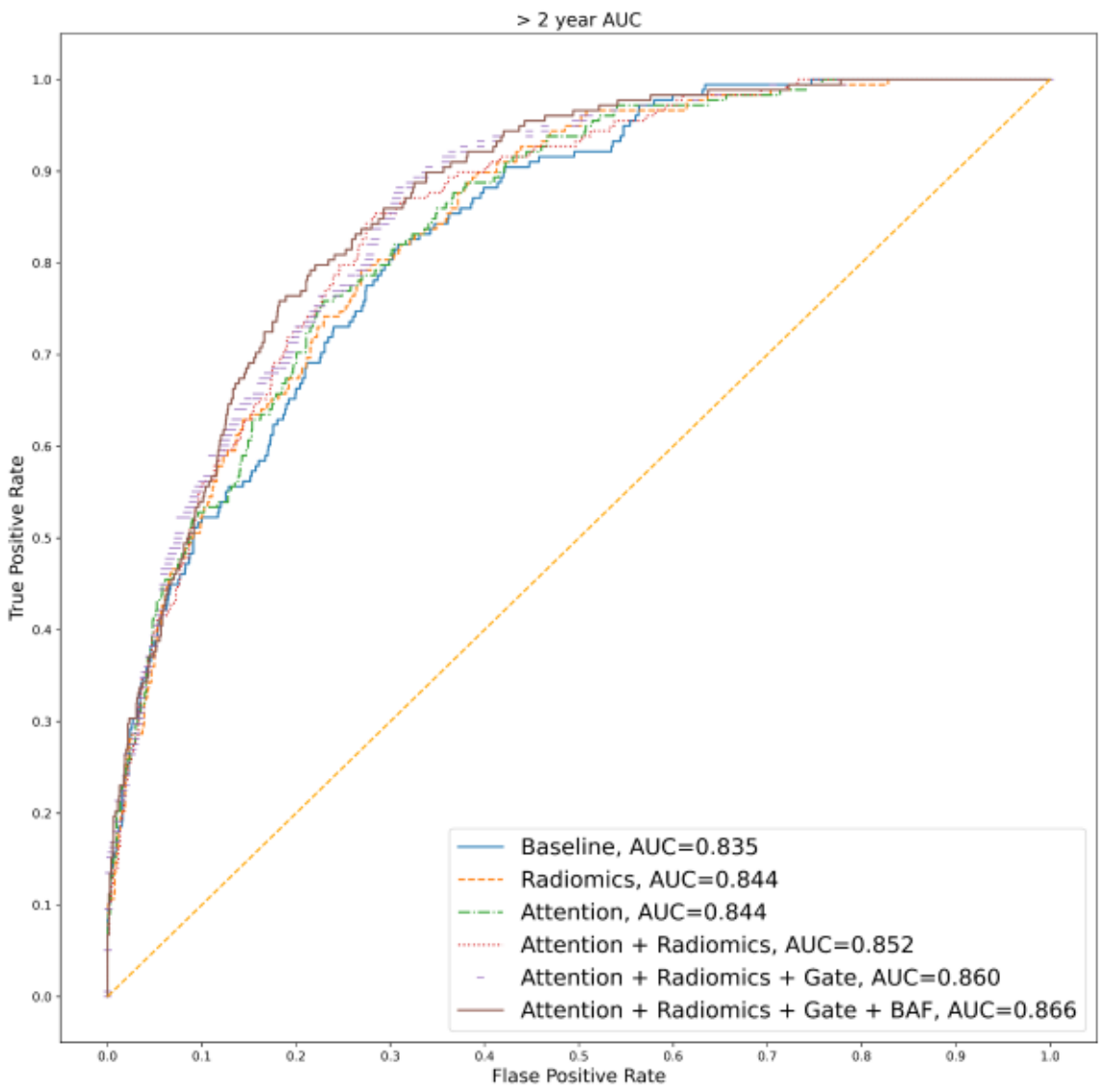

Fig. 7. ROC curves of ablation study results corresponding to 1-year, 2-year and >2-year AUC categories.

## 4. Discussion

In this paper, we report a new study with a number of unique characteristics. First, we introduce a deep learning model that can effectively handle sequential mammographic screenings using an attention mechanism that takes previous screenings into account. Sequential mammograms can indicate any abnormalities/changes in the breast tissue over time and the ability to detect these changes are paramount in a short-term risk prediction model. It is crucial to include previous screenings due to some cancers being more aggressive than others and the earlier cancer is detected, the higher the survivability of the patient.

With multiple mammograms combined to form a video input, we utilize a 3D self-attention method and 3D CNNs to handle such inputs. 3D CNNs differ from traditional 2D CNNs in that they can analyse spatial and temporal data simultaneously, making them well suited for analysing sequential mammograms. With our approach of inflating only certain convolutional kernels (as discussed in Section 2.3), pretrained off-the-shelf 2D CNN models can be converted to 3D CNNs with minimal increase in parameters. Our results show that our method of incorporating self-attention to detect global information across different screenings combined with 3D CNNs to extract local information in the spatiotemporal dimension can outperform state-of-the-art risk models that only use single screenings.

Our results show that having higher screening numbers produce better performance as more information is available with multiple screenings; however, using more screenings produce out-of-memory problems as observed in Table 6, whereby introducing Non-Local in Resnet layer 1 with just 2 sequential screenings already incurred this problem. Thus, the second novel contribution of our study is to introduce a new image based attention mechanism, called SHIFT (Spatial Channel Image Fastformer). Attention mechanisms including Non-Local attention introduced in [16] for video classification showed promising results in our previous studies for other organs/applications [22, 39]. Attention mechanisms can help weight the importance of different regions within the mammogram in predicting breast cancer risk. This is especially important in the case of multiple screenings and attention can help capture spatiotemporal relations between the mammograms. Understanding the importance of attention mechanisms, we looked to develop an attention mechanism that can handle inputs with high resolution (i.e., high resolution sequential mammograms) with high AUC results.

The Fastformer [21] is a transformer initially developed to be utilized in natural language processing with fast processing speed due to its linear complexity as compared to self-attention in the conventional transformer architecture [18], which has quadratic complexity and high memory costs. Fastformer has never been used on images and our study presents a new fast image based attention mechanism. In our study, we extended the original Fastformer paper [21] to incorporate channel attention (the original study only included spatial attention). We also performed some ablation studies on the original Fastformer architecture including the query addition, which was found to be redundant in our experiments. To the best of our knowledge, our study is the first to test the attention mechanism for breast cancer risk prediction. The results show that the inclusion of either Non-Local or SHIFT improves the deep learning architecture in risk prediction, with the latter producing better AUC results and much faster performance.

Third, we examined some new radiomics frequency based features generated from DCT and FFT transformations for breast cancer risk prediction. These new statistical features can capture additional information about the tissue characteristics relevant to cancer risk and they have been shown to work well for benign/malignant classification [27]. The results show that radiomic features complement deep learning features for risk prediction. By combining radiomic features and deep learning features, additional information can be captured on the texture and patterns of breast tissue and improve performance. Radiomic features and deep learning features may capture different aspects of the tissue, as radiomics extracts information on tissue texture and mammographic density [14, 40], while deep learning can learn additional patterns that are not easily extracted by traditional radiomic features. Our results show that combining both feature groups provides a more complete picture of the tissue patterns/characteristics. Previous studies tended to use either features independently [14], but the study results show that radiomic features provide supplementary information to deep learning features for better performance.

Fourth, our study introduces a new lightweight gating mechanism to handle deep learning features from both CC and MLO views. Both mammographic views provide relevant information pertaining to cancer risk prediction; however, the relative importance of each view may differ as certain views provide more information than others [9, 32, 35]. Thus, combining information from both views remains the best strategy. One common approach to combining features from multiple views is to use fully connected layers and different strategies devised in literature show promising results [32-34]. However, the number of parameters in the fully connected layers can become very large, which can lead to overfitting and making the model computationally expensive to train and evaluate. Our lightweight approach is easily implemented and is similar to the attention mechanism, whereby the model learns different weights/scaling factors that emphasize the importance of different views to optimally combine both views, thus improving overall performance. Similar to previous computer algorithm based studies [9, 32], our results showed that the CC view has a higher weightage than MLO in the final model.

Fifth, leveraging on results from our previous studies [15, 29, 32], we estimated bilateral asymmetry between left and right breasts and used this information to refine/finetune our model to boost its capabilities. Asymmetrical differences in appearance or density between left and right breasts indicate cancer development and machine learning models can be trained to detect/quantify these differences for risk prediction. We first introduced a method to refine our training dataset by removing controls that exhibited high bilateral asymmetry. To do this, we obtained an asymmetry risk score by generating risk scores for each independent view. We then filtered out controls in the training set that had high asymmetry scores, which could affect/confuse the training of the network. The results demonstrated that training the model on soft labels with the full dataset first and followed by training on the filtered dataset (also on soft labels), enhanced its performance. Training with soft labels can be beneficial as a regularization technique to reduce overfitting and to improve the generalization ability of the model.

Sixth, we examined our new methods on a comprehensive screening dataset consisting of 8,723 patients altogether with all 4 mammographic views (i.e., 34,892 mammograms), obtaining results that are better than state-of-the-art methods in the literature. As the CSAW dataset has multiple sequential screening mammograms, we can examine our new model’s performance in predicting short-term cancer risk, which is useful for prescribing personalized screening mammography procedures to individual women in this era of precision medicine [8, 15].

Although our study has promising results and novel observations, we acknowledge that this is a laboratory based retrospective data analysis study with several limitations that can be addressed in future studies. First, we had to duplicate screening mammograms that were missing in the CSAW dataset (due to patients not going for regular screening), which could have affected our model's performance. In future studies, we plan to evaluate our model's performance on a larger dataset(s) with each patient having consistent/regular screenings. Second, as our study was a developmental based study, no clinical testing on its utility or impact in assisting radiologists was evaluated. Third, cases in category 1 of our dataset consisted of patients diagnosed with cancer at <= 60 days from screening; these were presented as 1-year AUC results [8]. For a strict definition of risk prediction, these cases might be omitted, although the number of cases in the dataset will be considerably reduced, which we can consider examining in future studies with bigger datasets. Despite these limitations, our study presents a new deep learning architecture that utilizes attention, radiomics, multi-view gating and bilateral asymmetry based methods on a comprehensive screening dataset for short-term breast cancer risk prediction.

**Declaration of Competing Interests**

The authors of this manuscript declare that there are no competing interests and no relationships with any companies, whose products or services may be related to the subject matter of the article.

**Acknowledgement**
This work was supported by the Fundamental Research Grant Scheme (FRGS), Ministry of Higher Education Malaysia (MOHE), under grant FRGS/1/2022/ICT02/MUSM/02/1.

**Data availability**

The authors do not have permission to share data.

**Yeoh Hong Hui** is a PhD student at the School of Engineering, Monash University Malaysia. He is currently working on deep learning in medical imaging, primarily predicting breast cancer risk from mammograms.

**Andrea Liew** is enrolled as a PhD student at the School of Engineering, Monash University Malaysia. Her research interests include brain and breast lesion detection using deep learning techniques.

**Raphaël Phan** is Professor at Monash University, specializing in machine learning, security, cryptography, and malicious artificial intelligence. He has published over 90 journal papers and in excess of 120 conference papers.

**Fredrik Strand** is a Docent and MD PhD Radiologist within the Breast Imaging Unit at the Karolinska University Hospital. His research interests include application of new machine learning techniques to breast radiology images.

**Kartini Rahmat** is Professor and consultant radiologist at the University of Malaya Medical Centre. She is currently the principal investigator and grant holder for multiple breast oncology and cancer radiomics grants.

**Tuong Linh Nguyen** is a Research Fellow at the Centre for Epidemiology and Biostatistics in the School of Population Global Health. His current research lies in breast cancer.

**John L. Hopper** is a genetic epidemiologist and Professor at the University of Melbourne, where he is a Professorial Fellow and Director of the Centre for Epidemiology and Biostatistics in the School of Population Global Health.

**Maxine Tan** is a Senior Lecturer at the School of Engineering, Monash University. Her research interests include deep learning and quantitative imaging biomarkers for improving cancer screening, diagnosis, and prognosis assessment.